\documentclass{article}
\usepackage[scale={.83,.83}]{geometry}
\usepackage[latin1]{inputenc}
\usepackage{amsmath,amssymb,bbm,mathrsfs}
\usepackage{multicol}
\usepackage[hypertex]{hyperref}
\newlength{\textlength}
\newlength{\overlinelength}
\newcommand{\ovl}[2][.55]{\settowidth{\textlength}{$#2$}
  \setlength{\overlinelength}{0.1pt}
  \addtolength{\overlinelength}{0.75\textlength}
  \makebox[\textlength][s]{$#2$} \hspace{-.55\textlength}
  \hspace{-\overlinelength}\hspace{#1\overlinelength}
  \overline{\makebox[\overlinelength][s]{\vphantom{$#2$}}}
  \hspace{-#1\overlinelength}\hspace{.55\textlength}}

\newcommand{\VEV}[1]{\left\langle #1\right\rangle}

\begin{document}

\thispagestyle{empty}


\begin{center}
  {\large\bfseries Dimension-Five Operators in Grand Unified Theories}
  \\[6pt]
  {\bfseries Joshua Sayre, S\"oren Wiesenfeldt, and Scott Willenbrock}
  \\[6pt]
  {\slshape
    \begin{minipage}{.8\linewidth}
      \centering Department of Physics, University of Illinois at
      Urbana-Champaign,
      \\
      1110 West Green Street, Urbana, IL 61801, USA
    \end{minipage}
  }
\end{center}

\begin{abstract}
  \noindent
  Extensions of the standard model with low-energy supersymmetry
  generically allow baryon- and lepton-number violating operators of
  dimension four and five, yielding rapid proton decay.  The
  dimension-four operators are usually forbidden by matter parity.  We
  investigate to what extent the appearance of dimension-five
  operators at the Planck scale may be constrained by the different
  grand-unified gauge groups.  Dimension-five operators are suppressed
  in models based on $\text{E}_6$ and $\text{SU(3)}_C \times
  \text{SU(3)}_L \times \text{SU(3)}_R$, where four matter fields do
  not form a gauge singlet.  An intermediate scale offers the
  possibility to sufficiently suppress these dimension-five operators.
\end{abstract}

\begin{multicols}{2}

  \noindent
  Although the standard model (SM) is extremely successful, it is
  likely that it is only an effective theory, subsumed by a more
  fundamental theory at short distances.  If this more fundamental
  theory lies at a scale much greater than the weak scale,
  $M_\text{W}$, then weak-scale supersymmetry supplies a means to
  stabilize the hierarchy between these two scales.  Remarkably, with
  the renormalization group equations of the minimal supersymmetric
  extension of the standard model (MSSM) above the weak scale, the
  three gauge couplings (with $\alpha_1=\frac53\alpha^\prime$, such
  that hypercharge is a properly-normalized SU(5)-generator) meet at
  $M_\text{GUT} = 2\times 10^{16}$~GeV.  This supports the idea that
  the strong, weak, and electromagnetic interactions are unified into
  a grand unified theory (GUT) with a single gauge coupling.  It is
  striking that the experimental evidence for small but non-vanishing
  neutrino masses fits nicely in this framework, as $M_\text{GUT}$ is
  of the right order of magnitude to generate small Majorana masses
  for the neutrinos.

  Weak-scale supersymmetric models generically allow baryon- and
  lepton-number violating operators of dimension four: $Q L d^c$, $u^c
  d^c d^c$ and $L L e^c$.  Here $Q$ and $L$ denote the quark and
  lepton doublet superfields and $u^c$, $d^c$ and $e^c$ the up and
  down antiquark and the positron superfields, respectively.  Since
  these operators mediate very rapid proton decay, there must be an
  additional symmetry that forbids some or all of these operators.
  One possibility is matter parity, a ${\mathbb Z}_2$ symmetry, where
  the matter superfields are odd, and the Higgs and gauge superfields
  are even \cite{Dimopoulos:1981zb}.  Matter parity is a discrete
  subgroup of $\text{U(1)}_{B-L}$, and is equivalent to $R$ parity
  \cite{Martin:1992mq}.

  Even with matter parity, however, the baryon- and lepton-number
  violating dimension-five operators $QQQL$ and $u^c d^c u^c e^c$ are
  allowed.\footnote{These dimension-five operators can be forbidden by
    another discrete symmetry, baryon triality \cite{triality};
    however, this symmetry is not consistent with grand unification.}
  These operators are generated when color-triplet Higgs fields, which
  are generically present in GUTs, are integrated out. These Higgs
  fields must acquire masses of order $M_\text{GUT}$ in order to avoid
  rapid proton decay via these dimension-five operators.  With
  superpartners at the weak scale, the proton decay rate is typically
  near the experimental limit \cite{dim5-gut}.

  Regardless of grand unification, we should expect these
  dimension-five operators to be gravitationally induced
  \cite{dim5-grav}.  If they are suppressed by the Planck mass,
  $M_\text{P} = \left(8\pi G_N\right)^{-1/2} = 2\times 10^{18}$ GeV,
  their coefficients must be smaller than $10^{-7}$ in order to
  satisfy the bounds on proton decay.\footnote{Conversely, the decay
    rate is sufficiently suppressed if the scalar matter fields
    (sfermions) are as heavy as a few hundred TeV \cite{heavy-scalar}.
    However, in such scenarios fine-tuning is needed to leave one
    Higgs field at the weak scale.}  One way to explain these small
  coefficients is to argue that they are similar to Yukawa couplings
  \cite{Melfo:2003gk}; this idea may be realized in models with
  spontaneously-broken global \cite{flav} or gauged \cite{anomalous}
  flavor symmetries.  There is, however, considerable doubt that exact
  global symmetries exist in nature \cite{Witten:2000dt}.  Even if
  they do, it is not guaranteed that they can adequately suppress
  gravitationally-induced proton decay in the context of a
  grand-unified theory.  If the flavor symmetry is a gauge symmetry,
  it is very difficult to embed it in a grand-unified theory.

  In this paper we investigate to what extent the appearance of these
  operators at the Planck scale may be constrained by the different
  grand-unified gauge groups.  Among the most promising GUT groups are
  SO(10) \cite{so10}, $\text{E}_6$ \cite{Gursey:1975ki} and
  $\text{SU(3)}_C \times \text{SU(3)}_L \times \text{SU(3)}_R \equiv
  \text{G}_\text{TR}$ \cite{trinification}.  As rank-5 and rank-6
  groups, they have a $\text{U(1)}_{B-L}$ subgroup, and thus
  potentially preserve matter parity \cite{Martin:1992mq}.


  In SO(10), the MSSM matter fields, together with the right-handed
  neutrino, fit into the spinor representation, $\mathsf{16}_M$, while
  the MSSM Higgs fields are in the vector representation,
  $\mathsf{10}_H$.  Therefore SO(10) distinguishes between matter and
  Higgs fields.  The dimension-four operator
  $\left[\mathsf{16}_M\right]^3$ cannot appear, as it is not SO(10)
  invariant.  This is an improvement over the MSSM as well as SU(5),
  where the dimension-four operator $\ovl{\mathsf{5}}_M
  \ovl{\mathsf{5}}_M \mathsf{10}_M \ni Q L d^c + u^c d^c d^c + L L
  e^c$ is allowed by the gauge symmetry.

  Matter parity is automatic in SO(10) if one restricts the Higgs
  representations to be of even congruency class, so that the only
  operators allowed by the gauge symmetry contain an even number of
  matter fields \cite{Aulakh:1999cd}.  In particular, if
  $\text{U(1)}_{B-L}$ is broken by the vacuum expectation value (vev)
  of the SM singlet in $\ovl{\mathsf{126}}_H$, then matter parity
  emerges as a discrete gauge symmetry.  Right-handed neutrino masses
  are generated at the GUT scale via the renormalizable operator
  $\mathsf{16}_M \mathsf{16}_M \ovl{\mathsf{126}}_H$; this is
  desirable to generate eV-scale masses for active neutrinos via the
  see-saw mechanism.

  Alternatively, $\text{U(1)}_{B-L}$ can be broken by Higgs fields in
  the spinor representation.  Then the distinction between matter and
  Higgs fields is lost, with the undesirable consequence that matter
  parity is no longer automatic.  For example, consider the operator
  $\left[\mathsf{16}_M\right]^3 \mathsf{16}_H$, with a coefficient of
  order $1/M_\text{P}$.  This operator includes the product of SU(5)
  fields $\ovl{\mathsf{5}}_M \ovl{\mathsf{5}}_M \mathsf{10}_M N_H$.
  When the SM singlet field $N_H$ acquires its vev, $\VEV{N_H} \sim
  M_\text{GUT}$, the dimension-four operators $\ovl{\mathsf{5}}_M
  \ovl{\mathsf{5}}_M \mathsf{10}_M \ni Q L d^c + u^c d^c d^c + L L
  e^c$ are generated, suppressed only by $M_\text{GUT}/M_\text{P}$.
  Hence, we must impose matter parity (or some other symmetry) to
  forbid this operator.

  Whether matter parity is automatic or imposed, the dimension-five
  operator $\left[\mathsf{16}_M\right]^4 \ni QQQL + u^c d^c u^c e^c$
  is allowed by SO(10) symmetry.  We should expect this operator to be
  induced at the Planck scale, in which case its coefficient must be
  less than $10^{-7}$ to evade the bounds on proton decay.  Whether it
  is possible to arrange this with spontaneously-broken flavor
  symmetries is a model-dependent question.  For example, consider the
  rather complete SO(10) model of Albright and Barr
  \cite{Albright:2000dk}. This is an interesting model where a global
  flavor symmetry does not provide sufficient suppression of
  gravitationally-induced dimension-five operators.  The $\text{U(1)}
  \times \mathbbm{Z}_2 \times \mathbbm{Z}_2$ flavor symmetry allows
  the operator $\mathsf{16}_1 \mathsf{16}_2 \mathsf{16}_2
  \mathsf{16}_3 Y^\prime A$, where subscripts denote the generation.
  After $Y^\prime$ and $A$ acquire GUT-scale vevs, it generates the
  operator $Q_1 Q_2 Q_2 L_3$, which is suppressed by only
  \mbox{$\left(M_\text{GUT}/M_\text{P}\right)^2 \sim 10^{-4}$}.

  \smallskip


  Let us therefore proceed to $\text{E}_6$, where the MSSM matter
  fields are embedded in the fundamental representation,
  $\mathsf{27}_M$.  It decomposes with respect to
  $\text{SO(10)}\times\text{U(1)}$ as
  \begin{align} \label{eq:27dec} \mathsf{27} \to \mathsf{16}_1 +
    \mathsf{10}_{-2} + \mathsf{1}_4 \; .
  \end{align}
  In addition to the SO(10) spinor, which contains the MSSM matter
  fields, we find an SO(10) vector and a singlet, $S$.  Since
  \begin{align} \label{eq:27times27} \mathsf{27 \times 27} =
    \ovl{\mathsf{27}} + \ovl{\mathsf{351}} + \ovl{\mathsf{351}}^\prime
    \; ,
  \end{align}
  fermion masses can be generated via renormalizable Yukawa couplings
  to Higgs fields $\mathsf{27}_H$, $\mathsf{351}_H$, and
  $\mathsf{351}_H^\prime$ \cite{e6}, where
  \begin{alignat}{2}
    & \mathsf{351} & & \to \mathsf{10}_{-2} + \ovl{\mathsf{16}}_{-5} +
    \mathsf{16}_1 + \mathsf{45}_4 + \mathsf{120}_{-2} + \mathsf{144}_1
    \; , \label{eq:351}
    \\
    & \mathsf{351}^\prime & & \to \mathsf{1}_{-8} + \mathsf{10}_{-2} +
    \ovl{\mathsf{16}}_{-5} + \mathsf{54}_4 + \ovl{\mathsf{126}}_{-2} +
    \mathsf{144}_1 \; . \label{eq:351prime}
  \end{alignat}

  Contrary to SO(10), the product of three matter fields,
  $\left[\mathsf{27}_M\right]^3$, is gauge invariant.  Thus
  $\text{E}_6$ matter parity, under which $\mathsf{27}_M \to
  -\mathsf{27}_M$, is never automatic.  Despite this, $R$
  parity\footnote{To avoid confusion, we henceforth refer to matter
    parity as $R$ parity, to distinguish it from $\text{E}_6$ matter
    parity.}  may still be automatic, depending on which Higgs fields
  (and which vevs) are used to break $\text{E}_6$
  \cite{Martin:1992mq}.

  As in SO(10), $R$ parity is not conserved if the SM singlet field in
  the $\mathsf{16}_1$ component of $\mathsf{27}_H$, $N_H$, acquires a
  vev.  In this case, we generically have the two operators
  \begin{subequations} \label{eq:dim4}
    \begin{align} \label{eq:27^3} c \left[\mathsf{27}_M\right]^3 & \ni
      c \, \mathsf{16}_M \mathsf{16}_M \mathsf{10}_M
      \\
      \label{eq:27yukawa}
      y \,\mathsf{27}_M \mathsf{27}_M \mathsf{27}_H & \ni y \left(
        \mathsf{16}_M \mathsf{16}_M \mathsf{10}_H \right.
      \\
      & \mspace{42mu} + \left. \mathsf{16}_M \mathsf{10}_M
        \mathsf{16}_H + \mathsf{10}_M \mathsf{10}_M S_H \right) ,
      \nonumber
    \end{align}
  \end{subequations}
  where $y$ is the Yukawa coupling and we denote the SO(10) singlet
  Higgs field by $S_H$.  The first term in Eq.~(\ref{eq:27yukawa})
  generates quark and lepton masses when $\mathsf{10}_H$ acquires a
  weak-scale vev.  The last term generates a mass of order $y
  M_\text{GUT}$ for the matter fields in $\mathsf{10}_M$ when $S_H$
  acquires its vev, breaking $\text{E}_6$ to SO(10).  Integrating out
  the $\mathsf{10}_M$ field from Eqs.~(\ref{eq:dim4}), we generate the
  operator\footnote{Since the Yukawa coupling $y$ is a matrix in
    generation space, the coefficient of this operator should be read
    schematically.}
  \begin{align} \label{eq:integrate} 
    \frac{c\,y}{y M_\text{GUT}} \left[\mathsf{16}_M\right]^3
    \mathsf{16}_H
  \end{align}
  which yields the usual dimension-four baryon- and lepton-number
  violating operators when $N_H$ acquires its vev.  Therefore, we must
  impose $\text{E}_6$ matter parity to ensure that $R$ parity is
  respected.  Note that this operator is obtained regardless of the
  specific mechanism that gives mass to $\mathsf{10}_M$.  The only
  essential point of the above example is that $\text{U(1)}_{B-L}$ is
  broken by $N_H$.

  On the other hand, $R$ parity is automatic in $\text{E}_6$ if
  $\text{U(1)}_{B-L}$ is broken via the SM singlet Higgs field in the
  $\ovl{\mathsf{126}}_{-2}$ of $\mathsf{351}^\prime_H$
  \cite{Martin:1992mq}.  This is analogous to the case of SO(10),
  discussed above.  Now right-handed neutrino masses are generated via
  the renormalizable operator $\mathsf{27}_M \mathsf{27}_M
  \mathsf{351}^\prime_H \ni \mathsf{16}_M \mathsf{16}_M
  \ovl{\mathsf{126}}_H$.

  Whether or not $\text{E}_6$ matter parity is imposed, four matter
  fields, $\left[\mathsf{27}_M\right]^4$, do not form an $\text{E}_6$
  singlet.  This is an improvement upon SO(10), where
  $\left[\mathsf{16}_M\right]^4$ is allowed by the gauge symmetry.
  However, dimension-five operators can still be generated from
  higher-dimensional operators.  For example, consider the
  dimension-six operator $\left[\mathsf{27}_M\right]^4
  \ovl{\mathsf{27}}_H$, which contains the
  $\text{SO(10)}\times\text{U(1)}$ invariant operator
  $\left[\mathsf{16}_M\right]^4 \ovl{S}_H$.  We should expect this
  operator to be generated at the Planck scale with a coefficient of
  order $1/M_\text{P}^2$.  When $\ovl{S}_H$ acquires a vev, breaking
  $\text{E}_6$ to SO(10), it generates the usual SO(10) dimension-five
  operator, suppressed by only $M_\text{GUT}/M_\text{P}\sim 10^{-2}$.

  In the discussion thus far we have seen that the vevs of $S_H$ and
  $\ovl{S}_H$ generate a mass for $\mathsf{10}_M$ as desired
  [Eq.~(\ref{eq:27yukawa})], but also generate the unwanted
  dimension-five operators.  This is because both
  $\left[\mathsf{10}_{-2}\right]^2$ and $\left[\mathsf{16}_1\right]^4$
  contain an SO(10) singlet with $\text{U(1)}$-charges $\pm 4$.  If we
  replace $S_H$ with the $\mathsf{45}_4$ of $\mathsf{351}_H$ or
  $\mathsf{54}_4$ of $\mathsf{351}^\prime_H$, the same result is
  obtained.  Thus we see that {\em any Higgs field that gives mass to
    $\mathsf{10}_M$ also generates the dimension-five operators}.  In
  order to sufficiently suppress these operators, the vev of this
  Higgs field must be much smaller than $M_\text{GUT}$.  Hence, we
  must break $\text{E}_6$ at the GUT scale with Higgs fields that do
  not generate a mass for $\mathsf{10}_M$.  We then generate a mass
  for $\mathsf{10}_M$ at an intermediate scale,
  $M_\text{I}/M_\text{P}\le 10^{-7}$, such that the dimension-five
  operators are generated with a sufficiently-small coefficient.
  Since $\mathsf{10}_M$ consists of complete SU(5) multiplets, their
  presence at an intermediate scale does not upset gauge-coupling
  unification.

  Restricting ourselves to $\text{E}_6$ Higgs fields in the
  lowest-dimensional representations, namely $\mathsf{27}_H$,
  $\mathsf{78}_H$ (adjoint), $\mathsf{351}_H$, and
  $\mathsf{351}^\prime_H$, we have found a model that can successfully
  implement the above scenario.  Consider two Higgs fields in the
  $\mathsf{351}^\prime_H$ representation, $\mathsf{351}^\prime_{H1}$
  and $\mathsf{351}^\prime_{H2}$, and one field in the adjoint
  representation, $\mathsf{78}_H$.  The vev of the SO(10) singlet of
  $\mathsf{351}^\prime_{H1}$ breaks $\text{E}_6$ to SO(10), the
  $\ovl{\mathsf{126}}_{-2}$ component of $\mathsf{351}^\prime_{H2}$
  breaks $\text{U(1)}_{B-L}$, and the adjoint field completes the
  breaking to the SM.  The SM singlets, $S_M$ and $N_M$, obtain masses
  via couplings $\mathsf{27}_M \mathsf{27}_M \mathsf{351}^\prime_{H1}
  \ni \mathsf{1}_4 \mathsf{1}_4 \mathsf{1}_{-8}$ and $\mathsf{27}_M
  \mathsf{27}_M \mathsf{351}^\prime_{H2} \ni \mathsf{16}_1
  \mathsf{16}_1 \ovl{\mathsf{126}}_{-2}$, respectively.  It is
  noteworthy that this model automatically preserves $R$ parity; this
  was not required {\em a priori}.

  So far, $\text{E}_6$ is broken to the SM, and it is easy to check
  that we cannot generate the dangerous dimension-five operators:
  $\left[\mathsf{16}_1\right]^4$ has U(1) charge $+4$ and is a singlet
  under $\text{U(1)}_{B-L}$, whereas the singlet in
  $\mathsf{351}^\prime_{H1}$ has U(1) charge $-8$, and the vev of
  $\mathsf{351}^\prime_{H2}$ violates $\text{U(1)}_{B-L}$ by two
  units.  On the other hand, we have not yet generated a mass for the
  $\mathsf{10}_M$.  As discussed above, this can be accomplished by
  introducing another Higgs field in the $\mathsf{27}_H$,
  $\mathsf{351}_H$, or $\mathsf{351}^\prime_H$ representation,
  acquiring an intermediate-scale vev in the $S_H$, $\mathsf{45}_4$,
  or $\mathsf{54}_4$ direction, respectively.\footnote{It might be
    possible to use $\mathsf{351}^\prime_{H1}$ or
    $\mathsf{351}^\prime_{H2}$ for this purpose.}  This Higgs field
  also generates the unwanted dimension-five operators, but with a
  suppression of $M_\text{I}/M_\text{P}\le 10^{-7}$.

  In fact, this Higgs field might even break the electroweak symmetry,
  via a vev in the $\mathsf{10}_{-2}$ direction.  In this case, the
  mass hierarchy of the $\mathsf{10}_M$ is related to that of the SM
  fermions [see Eq.~(\ref{eq:27yukawa})].  For $M_\text{I}=10^{10}$
  GeV, we can expect the particles to be as heavy as $10^5$, $10^8$
  and $10^{10}$ GeV.  As mentioned above, the inclusion of three
  copies of SO(10)-fields {\sffamily 10} does not upset gauge-coupling
  unification and only slightly increases the value of $M_\text{GUT}$
  \cite{Kolda:1996ea}.  Moreover, it allows for perturbative
  unification: we effectively add $1.8$ copies at the weak scale,
  which increases the gauge coupling at $M_\text{GUT}$ by a factor of
  1.6. The mixing of $\mathsf{10}_M$ with down quarks and leptons is
  suppressed by $M_\text{W}/M_\text{I}$.

  It is beyond the scope of this work to construct a superpotential
  that generates the desired vevs.  It is a generic problem to explain
  the hierarchy between the weak and the GUT scale.  Whether the
  presence of the intermediate scale complicates this problem is an
  open question.  Note, for instance, that $M_\text{I}^2/M_\text{GUT}
  \sim M_\text{W}$.  Hence whatever solves the usual hierarchy problem
  may also address the existence of an intermediate scale.

  It is also possible that the intermediate scale is of the same order
  as the weak scale.  In such a scenario, the Yukawa interaction that
  gives rise to the $\mathsf{10}_M$ mass would have to be different
  from the one that gives rise to the mass of ordinary fermions, in
  order to avoid the first- and second-generation $\mathsf{10}_M$
  particles being unacceptably light.

  A Higgs field with an intermediate-scale vev could leave behind
  intermediate-scale particles that could disturb unification. In this
  scenario, however, it is the SO(10) singlet component that acquires
  the vev, so it is reasonable to expect that the particles acquire
  masses approximately within full SO(10) multiplets.  Then they do
  not upset gauge unification, as discussed above.

  A slight variation of this model is to use a second Higgs field in
  the adjoint representation instead of $\mathsf{351}^\prime_{H2}$.
  Since $\mathsf{78} \to \mathsf{1}_0 + \mathsf{45}_{0} +
  \mathsf{16}_{-3} + \ovl{\mathsf{16}}_{3}$, a vev in the
  $\mathsf{1}_0$ direction breaks $\text{E}_6$ to
  $\text{SO(10)}\times\text{U(1)}$.  Then, as above, a
  $\mathsf{351}^\prime_{H}$ vev in the $\ovl{\mathsf{126}}_{-2}$
  direction and a $\mathsf{78}_{H}$ vev in the adjoint direction
  complete the breaking to the SM, but with an extra unbroken U(1).
  This extra U(1) is broken at the intermediate scale where
  $\mathsf{10}_M$ and $S_M$ acquire mass; the mass of $S_M$ is
  generated via the non-renormalizable operator $\mathsf{27}_M
  \mathsf{27}_M \ovl{\mathsf{H}}\, \ovl{\mathsf{H}}$, where
  $\mathsf{H}=\mathsf{27}_H, \mathsf{351}_H, \mathsf{351}^\prime_H$,
  and is of order $M_\text{I}^2/M_\text{P}\sim 100$ GeV.

  \smallskip


  The results of $\text{E}_6$ also apply to the trinified group,
  $\text{SU(3)}_C \times \text{SU(3)}_L \times \text{SU(3)}_R \equiv
  \text{G}_\text{TR}$, where the equality of the three gauge couplings
  is enforced by a discrete symmetry, such as $\mathbbm{Z}_3$
  \cite{trinification,tr-models}.  It is a maximal subgroup of
  $\text{E}_6$ but is an interesting GUT candidate itself, in
  particular since it is broken to the SM without adjoint Higgs
  fields.

  Leptons, quarks and antiquarks are in different representations,
  \begin{align} \label{eq:trin-fields}
    \left(\mathsf{1,3},\ovl{\mathsf{3}}\right) +
    \left(\ovl{\mathsf{3}},\mathsf{1,3}\right) +
    \left(\mathsf{3},\ovl{\mathsf{3}},\mathsf{1}\right) \equiv
    \mathtt{L + Q}^c + \mathtt{Q} ,
  \end{align}
  which together complete the {\sffamily 27} of $\text{E}_6$.  Thus
  $\mathtt{Q}_M$ contains the quark doublet plus a color triplet, $B$,
  whereas the corresponding color anti-triplet, $B^c$, and the SM
  antiquark singlets are in $\mathtt{Q}^c_M$.  The lepton and
  antilepton fields are held by $\mathtt{L}_M$.

  In the minimal model, $\text{G}_\text{TR}$ is broken to the SM by a
  pair of Higgs fields, $\mathtt{L}_H$, when the SM singlets, $S_H$
  and $N_H$, acquire GUT-scale vevs.  These singlets are identical to
  the SO(10) singlet field and the SM singlet field in the
  $\mathsf{16}_1$-component of $\mathsf{27}_H$.  The vev of $N_H$
  breaks $R$ parity, and thus dimension-four baryon- and lepton-number
  violating operators are generated analogously to $\text{E}_6$, as
  discussed above [see Eqs.~(\ref{eq:dim4}) and (\ref{eq:integrate})].
  As in $\text{E}_6$, this can be evaded by imposing matter parity,
  under which $\mathtt{Q}_M$, $\mathtt{Q}^c_M$ and $\mathtt{L}_M$ are
  odd.  The vev of $S_H$ generates the dimension-five baryon-number
  violating operators via the dimension-six operator $\mathtt{Q}_M
  \mathtt{Q}_M \mathtt{Q}_M \mathtt{L}_M \ovl{\mathtt{L}}_H$, and
  hence is suppressed by only $M_\text{GUT}/M_\text{P}\sim 10^{-2}$,
  again analogous to $\text{E}_6$.

  We can, however, break $\text{G}_\text{TR}$ by a different pair of
  Higgs fields, $\Phi_a\left(1,\ovl{6},6\right)$.  This is the
  analogue of the $\text{E}_6$ model broken by two Higgs fields in the
  $\mathsf{351}^\prime_H$ representation.  Under $\text{SU(3)}_C
  \times \text{SU(2)}_L \times \text{SU(2)}_R \times \text{U(1)}_{B-L}
  \equiv \text{G}_{3221}$, the $\left(1,1,1\right)_0$ component of
  $\Phi_1$ acquires a vev, breaking $\text{G}_\text{TR}$ to
  $\text{G}_{3221}$. In doing so, the singlet $S_M$ acquires mass via
  the operator $\mathtt{L}_M\mathtt{L}_M\Phi_1$.  Next, the
  $\left(1,1,3\right)_{-2}$ component of $\Phi_2$ breaks
  $\text{G}_{3221}$ to the SM and the singlet $N_M$ acquires mass via
  the coupling $\mathtt{L}_M\mathtt{L}_M\Phi_2$.  This breaks
  $\text{U(1)}_{B-L}$ by two units, and hence $R$ parity is automatic.
  As in the $\text{E}_6$ scenario discussed above, the fields in
  $\mathsf{10}_M$ remain massless.  We must therefore introduce
  another Higgs field, $\mathtt{L}_H$, that acquires an
  intermediate-scale vev in the $S_H$ direction and generates a mass
  for $\mathsf{10}_M$ as well as the dimension-five operators
  suppressed by $M_\text{I}/M_\text{P}$.  Thus, as in $\text{E}_6$,
  the Higgs field that gives mass to $\mathsf{10}_M$ also generates
  the dimension-five operators.

  The trinified model offers another possibility not available in
  SO(10) and $\text{E}_6$.  We can remove the $\mathbbm{Z}_3$ that
  equates the three gauge couplings, and assume that some additional
  structure, such as string theory, is responsible for their equality.
  Then we can introduce a discrete symmetry under which quark and
  lepton fields transform differently.  As an example, consider a
  $\mathbbm{Z}_2$ symmetry, under which the quark fields are odd but
  the lepton fields even.  Then baryon-number violating interactions
  are absent altogether at the classical level, via operators of any
  dimension.  However, this discrete symmetry has mixed
  $\mathbbm{Z}_2-\text{SU(3)}_{L,R}$ anomalies, and may be strongly
  violated by quantum gravitational effects \cite{anomaly}.

  \smallskip


  In this paper we have shown that gravitationally-induced baryon- and
  lepton-number violating dimension-five operators are generically
  suppressed by $M_\text{GUT}/M_\text{P}\sim 10^{-2}$ in both
  $\text{E}_6$ and $\text{SU(3)}_C \times \text{SU(3)}_L \times
  \text{SU(3)}_R$, which violates the bounds on proton decay.
  However, if we allow for an intermediate scale, we may arrange for
  these operators to be suppressed by $M_\text{I}/M_\text{P}\le
  10^{-7}$, and thus evade these bounds.  The only models of this type
  that we found also automatically preserve $R$ parity, and thus
  forbid dimension-four baryon- and lepton-number violating operators.
  A superpotential that yields the discussed scenario remains to be
  constructed.

  \smallskip

  We are grateful for conversations and correspondence with
  C.~Albright, D.~Berenstein, H.~Dreiner, J.~Harvey, H.~M.~Lee,
  J.~Lykken, S.~Martin, and Y.~Shadmi.  S.~W. and S.~W. thank the
  Aspen Center for Physics and the DESY Theory Group for hospitality.
  This work was supported in part by the U.~S.~Department of Energy
  under contract No.~DE-FG02-91ER40677.


  {\small

  }
\end{multicols}
\end{document}